\documentclass{article}

\usepackage{microtype}
\usepackage{graphicx}
\usepackage{booktabs} 
\usepackage{caption}
\usepackage{subcaption}

\usepackage{hyperref}

\usepackage[accepted]{icml2019}

\icmltitlerunning{Visualizing and Understanding Self-attention based Music Tagging}

\begin{document}

\twocolumn[
\icmltitle{Visualizing and Understanding Self-attention based Music Tagging}



\icmlsetsymbol{equal}{*}

\begin{icmlauthorlist}
\icmlauthor{Minz Won}{mtg}
\icmlauthor{Sanghyuk Chun}{clova}
\icmlauthor{Xavier Serra}{mtg}
\end{icmlauthorlist}

\icmlaffiliation{mtg}{Music Technology Group, Universitat Pompeu Fabra, Barcelona, Spain}
\icmlaffiliation{clova}{Clova AI Research, Naver Corp., Seongnam, Korea}

\icmlcorrespondingauthor{Minz Won}{minz.won@upf.edu}

\icmlkeywords{Machine Learning, ICML}

\vskip 0.3in
]



\printAffiliationsAndNotice{}  

\begin{abstract}
Recently, we proposed a self-attention based music tagging model \cite{won2019attention}. Different from most of the conventional deep architectures in music information retrieval, which use stacked $3\times3$ filters by treating music spectrograms as images, the proposed self-attention based model attempted to regard music as a temporal sequence of individual audio events. Not only the performance, but it could also facilitate better interpretability. In this paper, we mainly focus on visualizing and understanding the proposed self-attention based music tagging model.

\end{abstract}

\section{Introduction}
\label{introduction}
As deep learning based research successfully demonstrated its versatility, interest in interpretability of deep learning models has been increased together. In the field of computer vision (CV), researchers tried to interpret the reasons and mechanisms behind the predictions of deep architectures by: mapping intermediate feature activities to the input space using deconvolution \cite{zeiler2014visualizing}, highlighting class discriminative localization map using gradients of the target class (Grad-CAM) \cite{selvaraju2017grad}, and perturbing input to determine local importance (LIME) \cite{ribeiro2016should}. Motivated by the previous works, music information retrieval (MIR) researchers also endeavored to understand their deep architectures. Visualization using deconvolution was demonstrated with a genre classification model \cite{choi2016explaining} and an instrument recognition model \cite{han2017deep}. Especially, Choi et al. proposed an auralization method to interpret the network by listening to the reconstructed signal from deconvolved spectrograms \yrcite{choi2016explaining}. To explain the predictions of a singing voice detection model, Mishira et al. adopted LIME \yrcite{mishra2017local} and feature inversion \yrcite{mishra2018understanding}.

However, the aforementioned models for MIR are yet less interpretable. All of the introduced models are using stacked $3\times3$ filters, which was originally designed for the image processing, on spectrogram inputs. This architecture captures spectro-temporal local features in each layer, while music is a temporal sequence of individual audio events. From this motivation, we recently proposed a new architecture design for music tagging that captures timbral local features using CNN and learns their temporal relation using self-attention mechanism \cite{won2019attention}. The proposed model is not only good in its performance but also facilitates interpretable visualization. In following sections, we summarize the model architecture (Section \ref{model}), visualize learned information (Section \ref{visualization}), and finalize the paper by discussing future work (Section \ref{futurework}).

\section{Model Architecture}
\label{model}
The self-attention based music tagging model \cite{won2019attention} consists of two parts: front-end that captures timbral local information and back-end that models temporal relation of the captured local features. Although two different front-ends were introduced in the paper, in this work, we only use spectrogram-based CNN which is equivalent to the front-end of Pons et al. \yrcite{pons2017end}. The front-end CNN consists of vertical (i.e. $86\times7$) and horizontal (i.e. $1\times129$) filters.

The back-end of the model is identical to the encoder of the Transformer \cite{vaswani2017attention}. It consists of stacks of self-attention layers and was originally designed to solve natural language processing tasks. A self-attention module computes the response at a location in a sequence by attending to all locations within the same sequence. Since music composes its semantics based on the relations between components in sparse positions, we can adopt self-attention layers for music sequence modeling. Although features from the front-end are not discrete like language, self-attention has already been successfully plugged into computer vision architectures \cite{wang2018non} which use linear features. 

In this work, we used stacks of 2 multi-head attention layers with 8 attention heads which showed the best performance for $\approx$4.1s input sequence. Implementation details are available online \footnote{https://github.com/minzwon/self-attention-music-tagging} for reproducibility.

\begin{figure}[ht!]
    \centering
    \begin{subfigure}[h]{0.48\linewidth}
        \centering
        \includegraphics[width=\linewidth]{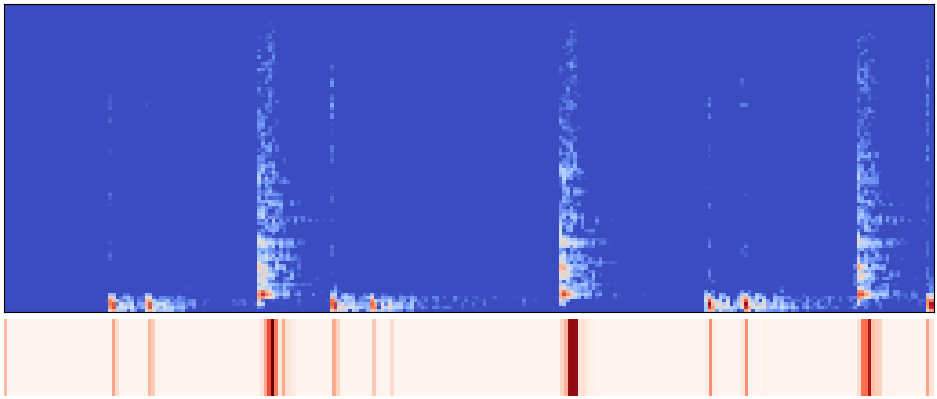}
        \caption{Tag - Drums}
        \label{fig:attention_heatmap_a}
    \end{subfigure}
    \begin{subfigure}[h]{0.48\linewidth}
        \centering
        \includegraphics[width=\linewidth]{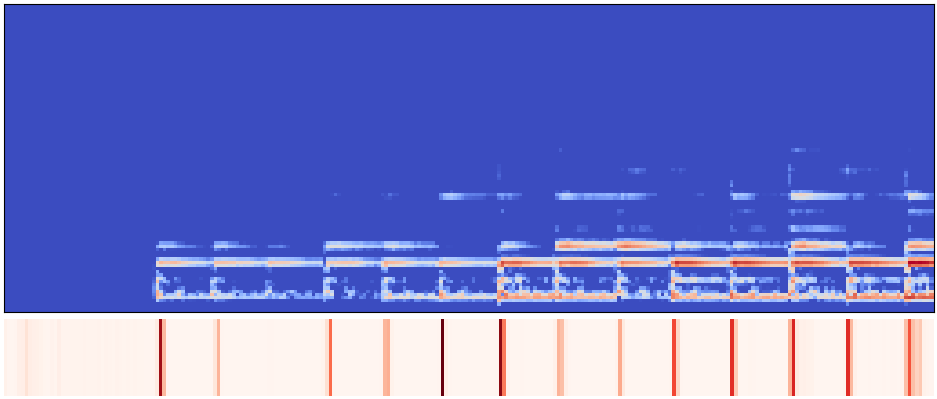}
        \caption{Tag - Piano}
        \label{fig:attention_heatmap_b}
    \end{subfigure}
    \begin{subfigure}[h]{0.48\linewidth}
        \centering
        \includegraphics[width=\linewidth]{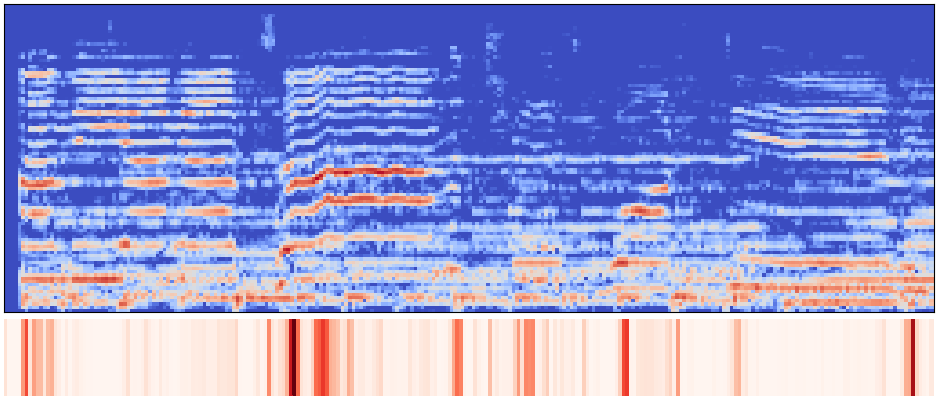}
        \caption{Tag - Vocal}
        \label{fig:attention_heatmap_c}
    \end{subfigure}
    \begin{subfigure}[h]{0.48\linewidth}
        \centering
        \includegraphics[width=\linewidth]{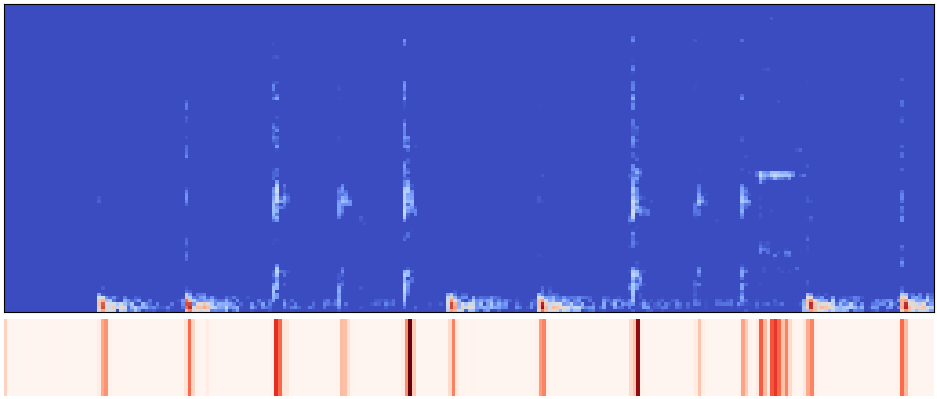}
        \caption{Tag - No Vocal}
        \label{fig:attention_heatmap_d}
    \end{subfigure}
    \begin{subfigure}[h]{0.48\linewidth}
        \centering
        \includegraphics[width=\linewidth]{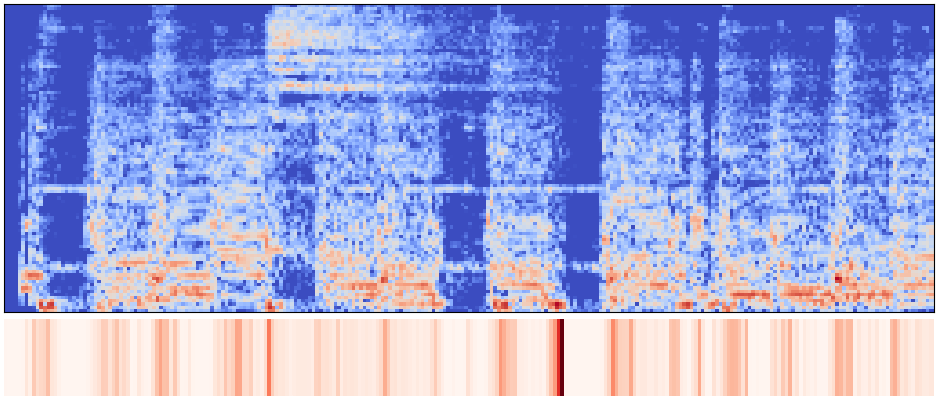}
        \caption{Tag - Loud}
        \label{fig:attention_heatmap_e}
    \end{subfigure}
    \begin{subfigure}[h]{0.48\linewidth}
        \centering
        \includegraphics[width=\linewidth]{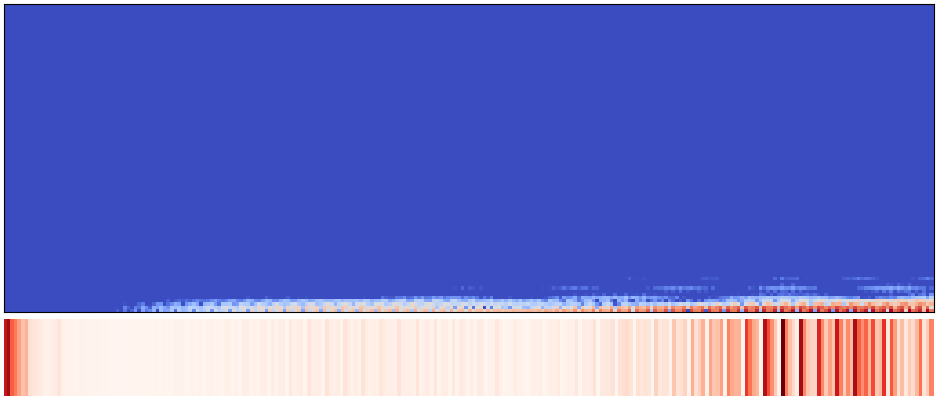}
        \caption{Tag - Quiet}
        \label{fig:attention_heatmap_f}
    \end{subfigure}
    \begin{subfigure}[h]{0.48\linewidth}
        \centering
        \includegraphics[width=\linewidth]{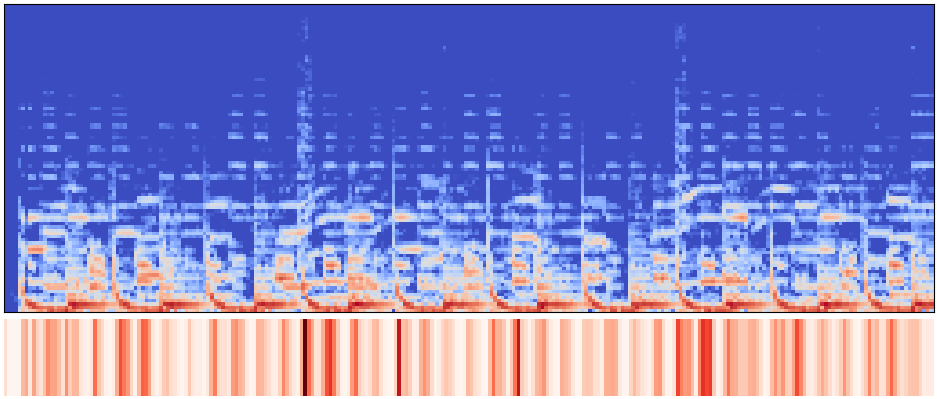}
        \caption{Tag - Techno}
        \label{fig:attention_heatmap_g}
    \end{subfigure}
    \begin{subfigure}[h]{0.48\linewidth}
        \centering
        \includegraphics[width=\linewidth]{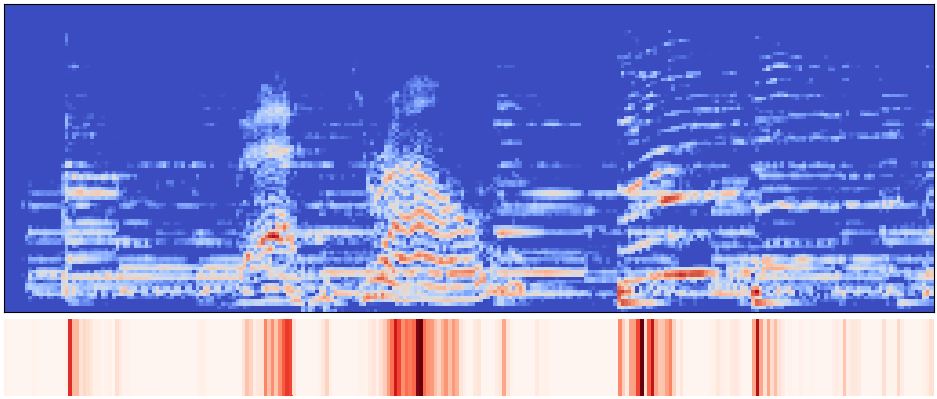}
        \caption{Tag - Country}
        \label{fig:attention_heatmap_h}
    \end{subfigure}

    \caption{Attention heat maps.}
    \label{fig:attention_heatmap}
\end{figure}

\section{Visualization}
\label{visualization}
\noindent\textbf{Attention Heat Map.}
To understand the behavior of the model, it is important to know which part of the audio the machine pays more attention to. To this end, we summed up attention scores from each attention head and visualized them. Figure \ref{fig:attention_heatmap} shows log mel-spectrograms and their attention heat maps. For simplification, we only visualized the attention heat map of the last attention layer. The model tends to pay more attention to more informative parts for music tagging. However, as shown in Figure \ref{fig:attention_heatmap_d} and \ref{fig:attention_heatmap_f}, attention heat maps always highlight parts with more energy although they were tagged as \textit{no vocal} and \textit{quiet}, respectively. Also, it is difficult to interpret the reason of tagging if the tags are related to longer-term information (Figure \ref{fig:attention_heatmap_e} and \ref{fig:attention_heatmap_f}). Attention heat maps can pinpoint where the machine pays attention, but they cannot provide reasons for the classification or tagging.

\begin{figure}[ht!]
    \centering
    \begin{subfigure}[h]{0.48\linewidth}
        \includegraphics[width=\linewidth]{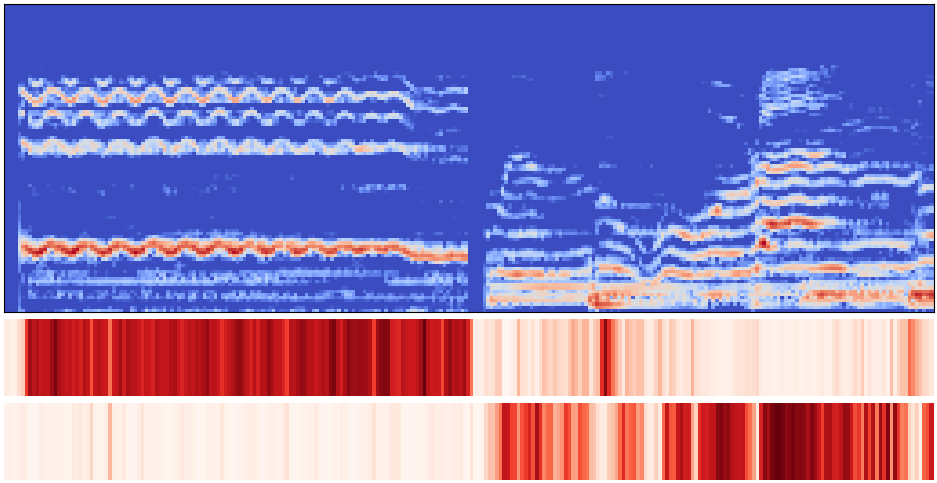}
        \caption{Female + Male}
        \label{fig:tagwise_contrib_a}
    \end{subfigure}
    \begin{subfigure}[h]{0.48\linewidth}
        \includegraphics[width=\linewidth]{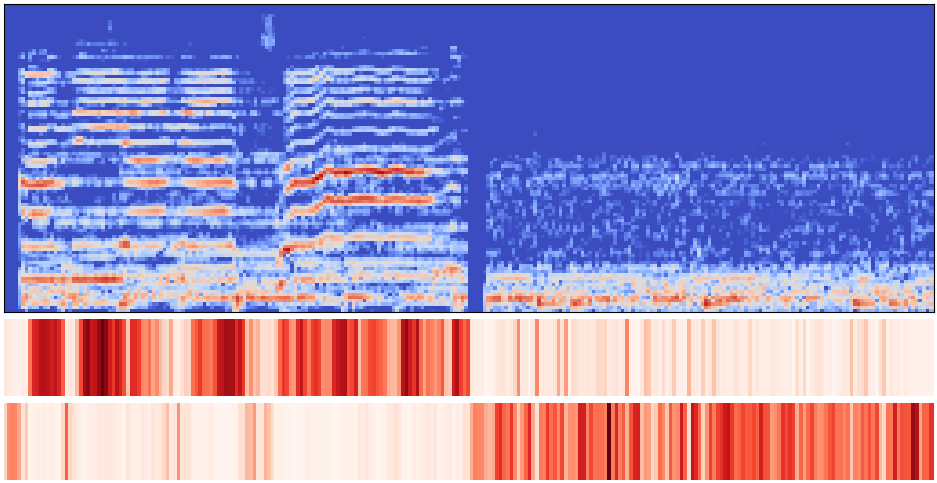}
        \caption{Vocal + No Vocals}
        \label{fig:tagwise_contrib_b}
    \end{subfigure}
    \begin{subfigure}[h]{0.48\linewidth}
        \includegraphics[width=\linewidth]{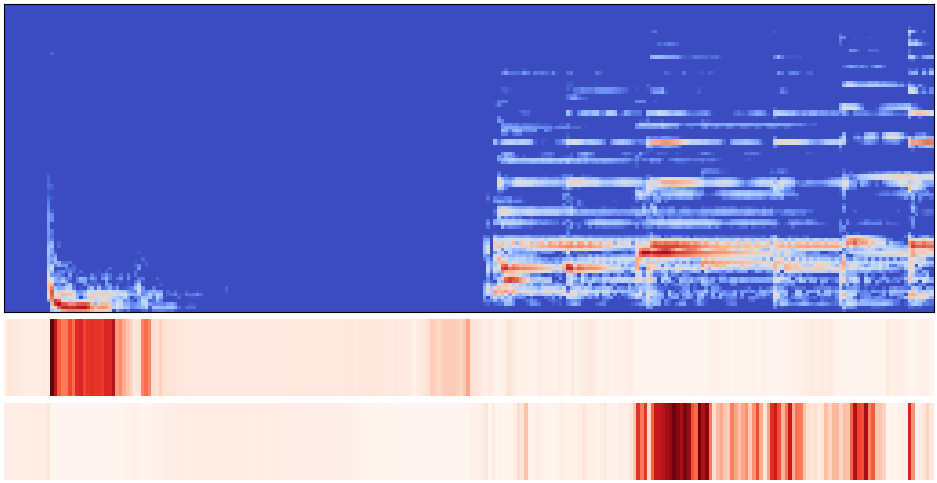}
        \caption{Drums + Harp}
        \label{fig:tagwise_contrib_c}
    \end{subfigure}
    \begin{subfigure}[h]{0.48\linewidth}
        \centering
        \includegraphics[width=\linewidth]{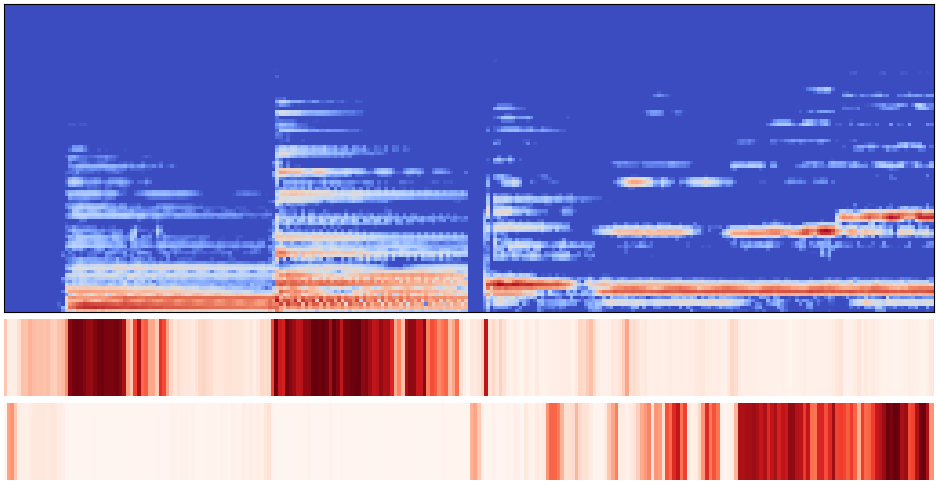}
        \caption{Piano + Flute}
        \label{fig:tagwise_contrib_d}
    \end{subfigure}
    \begin{subfigure}[h]{0.48\linewidth}
        \centering
        \includegraphics[width=\linewidth]{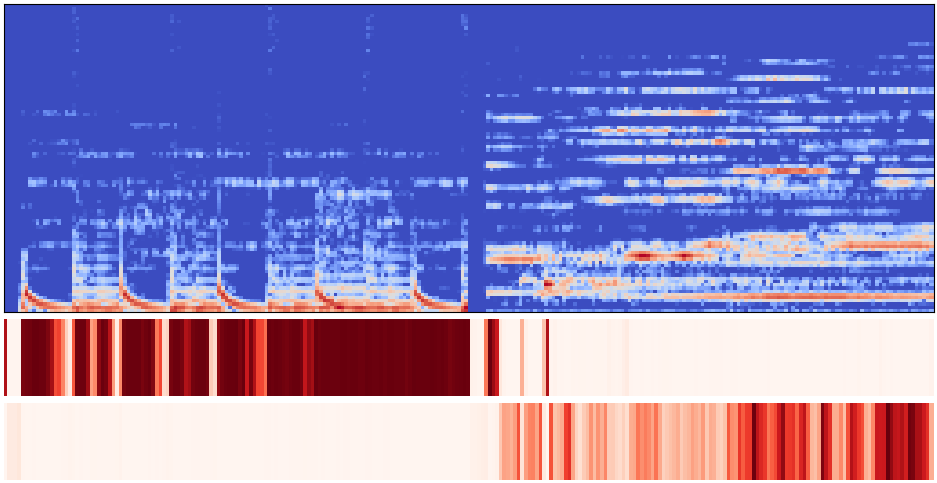}
        \caption{Techno + Classic}
        \label{fig:tagwise_contrib_e}
    \end{subfigure}
    \begin{subfigure}[h]{0.48\linewidth}
        \includegraphics[width=\linewidth]{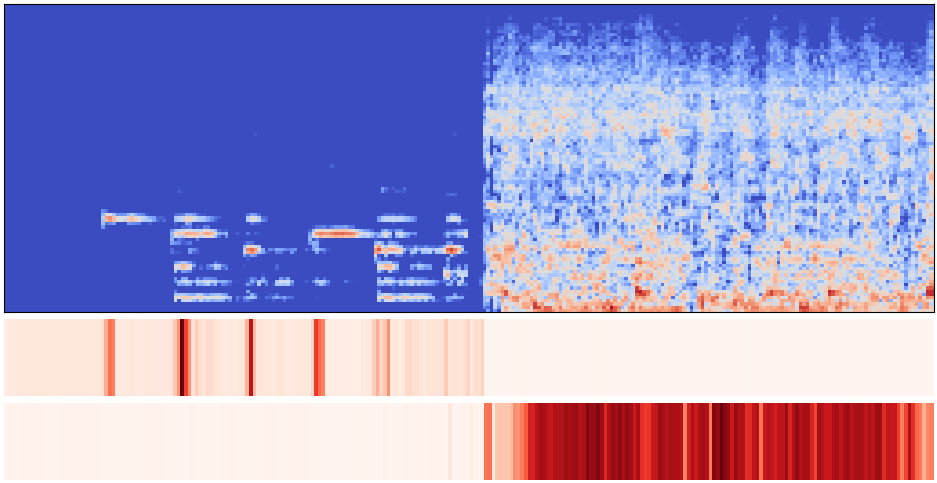}
        \caption{Classic + Metal}
        \label{fig:tagwise_contrib_f}
    \end{subfigure}

    \begin{subfigure}[h]{0.48\linewidth}
        \includegraphics[width=\linewidth]{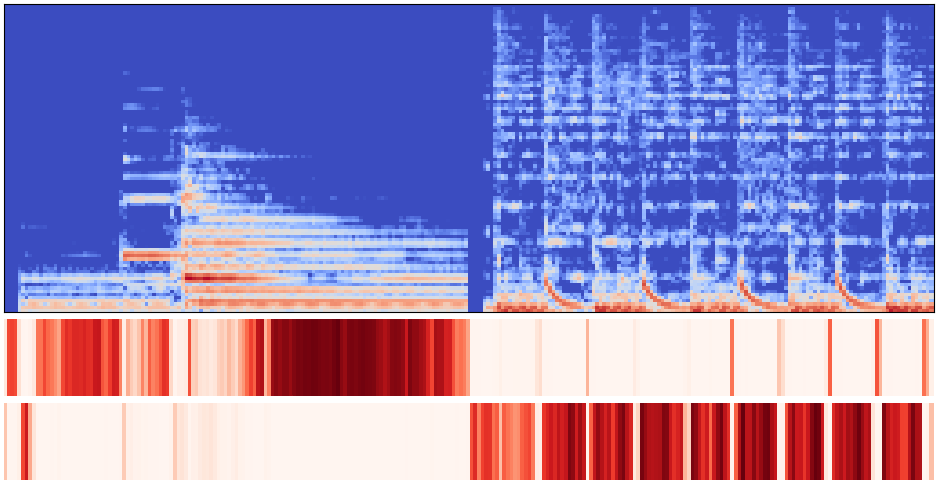}
        \caption{Slow + Fast}
        \label{fig:tagwise_contrib_g}
    \end{subfigure}
    \begin{subfigure}[h]{0.48\linewidth}
        \centering
        \includegraphics[width=\linewidth]{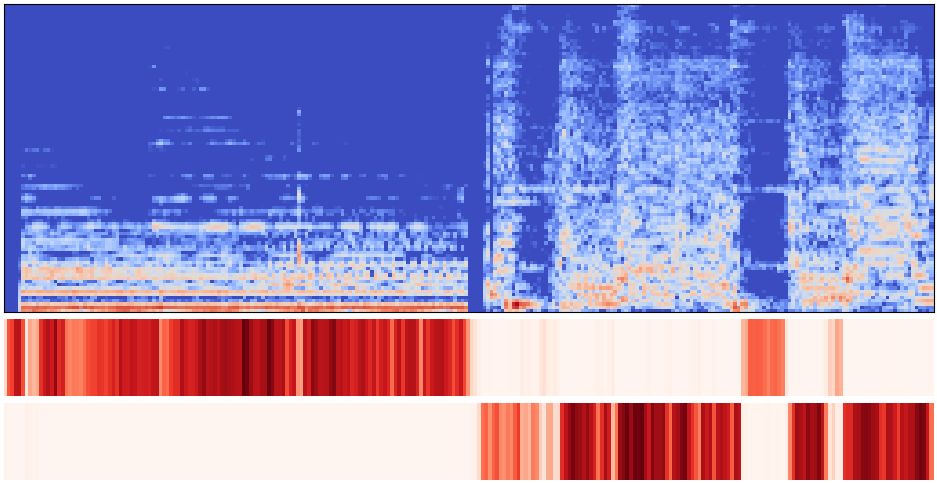}
        \caption{Quiet + Loud}
        \label{fig:tagwise_contrib_h}
    \end{subfigure}
    \caption{Tag-wise contribution heat maps.}
    \label{fig:tagwise_contrib}
\end{figure}

\noindent\textbf{Tag-wise Contribution Heat Map.}
To emphasize which part of the audio is more relevant to each tag, we visualized tag-wise contribution heat maps (Figure \ref{fig:tagwise_contrib}). We manually changed the attention score of the last attention layer. For each time step, we manipulated the attention score as 1 and set scores of other parts as 0 so that we can see the contribution of each time bin to each tag. This tag-wise contribution heat map is inspired by the manual attention weight adjustment proposed by Lee et al. \yrcite{lee2017interactive}. To compare the different contribution of different audio, we concatenated two spectrograms and fed them through the network. For example, Figure \ref{fig:tagwise_contrib_a} shows a concatenated spectrogram of \textit{female} and \textit{male} voice. The first row heat map highlights contribution to the \textit{female} tag and the second row indicates contribution to the \textit{male} tag. We repeated this for instruments (\ref{fig:tagwise_contrib_a}, \ref{fig:tagwise_contrib_b}, \ref{fig:tagwise_contrib_c}, \ref{fig:tagwise_contrib_d}), genres (\ref{fig:tagwise_contrib_e}, \ref{fig:tagwise_contrib_f}), and moods/themes (\ref{fig:tagwise_contrib_g}, \ref{fig:tagwise_contrib_h}). Different from the attention heat map, the tag-wise contribution heat map can facilitate better interpretation of the classification since it visualizes contribution to each tag. By comparing Figure \ref{fig:attention_heatmap_d} and \ref{fig:tagwise_contrib_b}, we could figure out that the model pays more attention to parts with more energy but the contribution to \textit{no vocal} tag is different from the attention heat map. Interestingly, the second half of the spectrogram in Figure \ref{fig:tagwise_contrib_h} has a temporary silence and the contribution heat map for \textit{quiet} shows an according short activation.


\section{Future Work}
\label{futurework}
We demonstrated the interpretability of our proposed model with a use case of music tagging. Since the architecture design is not task-specific, it can be applied to solve general MIR problems. However, the front-end of the proposed model is yet less interpretable. We could highlight relevant parts of audio for each tag, but still don't know which frequency band or timbral information is important for each tag. By adopting gradient-based visualization methods in the front-end, one can expect better interpretability.

\section*{Acknowledgements}

This work was funded by the predoctoral grant MDM-2015-0502-17-2 from the Spanish Ministry of Economy and Competitiveness linked to the Maria de Maeztu Units of Excellence Programme (MDM-2015-0502).

\nocite{langley00}

\bibliography{example_paper}
\bibliographystyle{icml2019}

\end{document}